# 3D Object Imaging through Scattering Media

Xiangsheng Xie[1, 2], Huichang Zhuang[2], Hexiang He[3, *], Xiaoqing Xu[2], Haowen Liang[2], Yikun Liu[2] and Jianying Zhou[2, *]

[1]Department of Physics, College of Science, Shantou University, Shantou, Guangdong 515063, China

[2]State Key Laboratory of Optoelectronic Materials and Technologies, Sun Yat-sen University, Guangzhou 510275, China

[3] School of Physics and Optoelectronic Engineering, Foshan University, Foshan 528000, China

[*]**Corresponding authors: sysuhhx@163.com, stszjy@mail.sysu.edu.cn**

**Abstract:** Human ability to visualize an image is usually hindered by optical scattering. Recent extensive studies have promoted imaging technique through turbid materials to a reality where color image can be restored behind scattering media in real time. The big challenge now is to recover a 3D object in a large field of view with depth resolving ability. Here, we reveal a new physical relationship between speckles generated from objects at different planes. With a single given point spread function, 3D imaging through scattering media is achieved even beyond the depth of field (DOF). Experimental testing of standard scattering media shows that the original DOF can be extended up to 5 times and the physical mechanism is depicted. This extended 3D imaging is expected to have important applications in science, technology, bio-medical, security and defense.

## Introduction

Light scattering is an obstacle for optical imaging when our sights are blocked by smoke, mist, anisotropy biological tissues, or air turbulence. As science and technique progress, we have witnessed steady development for optical imaging recovery through scattering media. Among various novel technologies, guide star assisted wavefront shaping[1-5], holography[6], scattering matrix measurement[7] and speckle correlation[2,8] are the most effective. However, these technologies are valid mainly for detecting the planer shape of the hidden object, where its depth and location information is rarely retrieved. The depth information, or 3-dimensional (3D) imaging, is vital for a wide range of real-world applications. For example, the exact location of the tumor and its axial structure should be determined before any therapy is operated. Furthermore, added axial information will greatly assist the determination of the nature of an object. To achieve this objective, a great deal of effort has been made and new techniques have been introduced in recent years. Velten et al. developed an ultrafast time-of-flight imaging method for recovering 3D shape around a corner[9]. Singh et al. made use of digital holography technique and successfully recovered a 3D object through a thin scattering medium[10]. Takasaki et al.[11] and Liu et al.[12] demonstrated 3D imaging behind diffusers by phase-space analysis methods. Wavefront shaping technique can render 3D images by taking advantage of temporal property of photo-acoustic tag[13,14]. "Memory effect" (ME) of a thin scattering medium was applied in axial direction to obtain 3D images with the help of a fluorescence or a nonlinear guide star[15,16]. Speckle correlation technique exploits the lens-like imaging ability of a thin scattering medium[2]. By introducing a reference point source and adjusting the imaging planes, Singh et al. obtained depth resolved images behind a scattering medium[17]. Among these improved techniques, the speckle correlation method appears very promising, as it does not need any coherent sources, raster-scanning mechanism, nor any time-consuming wavefront shaping devices. As reported in our last paper[4], a well-designed optical system generates a Point-Spread-Function (PSF) which is helpful for high speed color imaging through scattering media with a large Field-Of-View (FOV). However, similar to most of the imaging techniques based on ME, they commonly suffer from a limited Depth-Of-Focus (DOF) in axial detection. One straightforward solution is capturing image slices in different axial positions supplemented with a series of corresponding PSFs. However, determination of PSFs for different objective planes complicates the system hence recovery of 3D image appears low and less practical. It is thus fundamentally important to study the physical relationship between the depth of the object and its speckle properties.

Here, we reveal a new physical relationship between the speckle properties in the far field and PSFs from different object planes after a thin scattering medium. Thus a single PSF is effectively applied to quantitatively retrieve 3D imaging through the scattering media. Surprisingly, the range of $z$-axial can be retrieved even beyond the original DOF. Prove-of-concept is demonstrated both with simulation and with experiment that the retrieved imaging is shown with good quality and high fidelity. Experimental results show a 5 times extension of *z*-axis range with a single PSF compared with original DOF. A rigorous theoretical analysis is presented to support the numerical simulation results.

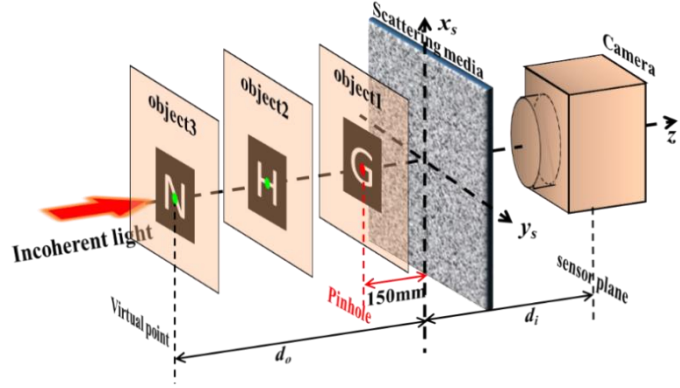

**Figure 1.** Schematic of deconvolution 3D imaging beyond DOF limit through a scattering medium. Virtual PSFs from virtual point (green) can be calculated with PSF from a real pinhole (red).

## Principle

For an imaging system, a point at the object plane generates a nearly identical pattern on the image plane (PSF), which represents the properties of the imaging system[18, 19]. The intensity distribution of the light field under incoherent illumination is described with a convolution, $I=O*PSF$, where $I$ and $O$ are intensity distribution on the image plane and the object plane respectively, the symbol * denotes a convolution operation.

When a thin scattering medium is introduced into the imaging system, the *PSF* becomes a speckle pattern determined by the property of the scattering medium. This scattering pattern is shown to be shift-invariant within the ME range[20-22]. The imaging process can be denoted as a correlation function of the *PSF* and the object distribution function[4, 20]:

$$I(x_i, y_i) = F(kx_i/d_i, ky_i/d_i) \iint PSF(x_i, y_i; x_o, y_o) O(x_o, y_o) dx_o dy_o, \qquad (1)$$

where *k* is the wavevector, the subscripts *i* and *o* denote the image and the object plane respectively. $F(kx_i/d_i, ky_i/d_i)$ is defined as a form factor function which acts as a field stop placed in the focal plane to limit the observed *FOV*. The convolution equation can be rewritten in the form of *I=F•(O*PSF)*. Although the *PSF* in this case is very complicated, the deconvolution process[4, 23] can restore *O* by considering the imaging system as a "black box". Obviously, deconvoluted imaging is valid only if the reference point and the object are on the same plane (focus). The image will become blurred or even invisible as long as the object deviates from the reference point (Fig.2 (f) of Ref. 4). The deviated distance between which the image can be restored is determined by the DOF.

For better understanding the relationship between different PSFs, as shown in Fig. 1, the complicated *PSF* is calculated by treating the thin scattering medium as a random phase mask[24] $TM(x_s, y_s)$. The light field distribution on the imaging plane is[20],

$$h(x_i, y_i; x_o, y_o) = Fr\{TM \cdot Fr\{\delta(x_o, y_o)\}_{do}\}_{di}, \qquad (2)$$

where $Fr\{\}_{di}$ and $Fr\{\}_{do}$ represent Fresnel diffraction with distance of $d_i$ and $d_o$ respectively. The intensity of *PSF* under incoherent illumination is recorded, which can be written as the square of the module of the light field: $PSF(x_i, y_i; x_o, y_o) = |h(x_i, y_i; x_o, y_o)|^2$.

By applying the autocorrelation theorem and rewriting equation with a form of Fourier transform. The *PSF* becomes,

$$PSF(x_i, y_i; x_o, y_o) = \mathscr{F}\left\{\mathscr{F}_a^{-1}\{A(\alpha, \beta)\}_{\alpha \to \frac{x_s}{2\lambda f}, \beta \to \frac{y_s}{2\lambda f}}\right\}_{x_s \to \frac{x_i}{\lambda d_i}, y_s \to \frac{y_i}{\lambda d_i}}, \qquad (3)$$

where $A(\alpha, \beta) = \sqrt{1/4i\lambda f} e^{-\frac{ik}{8f}(\alpha^2+\beta^2)} * [TM(\alpha, \beta)TM^*(\alpha - x_s, \beta - y_s)]$, $\mathscr{F}\{\}$ is a symbol of Fourier transformation, *f* is the focal length of the "scattering lens" which meets the formula of object-image distance ($1/f=1/d_i+1/d_o$).

When the object distance $d_o$ is translated to $d'_o$, the new PSF becomes,

$$PSF'(x_i, y_i; x_o, y_o) = \mathscr{F}\left\{\mathscr{F}_a^{-1}\{A(\alpha,\beta)\}_{\alpha \to \frac{x_s}{2\lambda f'}, \beta \to \frac{y_s}{2\lambda f'}}\right\}_{x_s \to \frac{x_i}{\lambda d_i}, y_s \to \frac{y_i}{\lambda d_i}}, \quad (4)$$

where the transform variables in $\mathscr{F}_a^{-1}$ is changing to $\alpha \to \frac{x_1}{2\lambda f'}$ and $\beta \to \frac{y_1}{2\lambda f'}$, $1/f' = 1/d_i + 1/d'_o$.

The new *PSF* is equal to the original one except the scaling in transform variable,

$$PSF'(x_i, y_i) = m^2 PSF(mx_i, my_i). \quad (5)$$

*m* is the scaling factor, and it is written by

$$m = \frac{f}{f'} = \frac{(d_i + d_o)d'_o}{(d_i + d'_o)d_o}. \quad (6)$$

A more detailed derivation of equation (2)-(6) can be found in the Supplementary Information.

Based on equation (5), the difference between PSFs in various object planes is only on the size scale. It turns out to be an interesting and important conclusion: once a PSF is obtained, virtual PSFs of other object planes can be calculated by resizing the PSF with a scaling factor *m*.

**Prove-of-concept simulation**

In an attempt to verify the above theory, a numerical simulation is implemented without using the above theoretical result. As shown in Fig. 1, the propagation of the light field from a pinhole through a scattering medium to the imaging plane is commonly described as two steps of Fresnel diffraction. The first step is Fresnel diffraction from the pinhole to the plane in front of the scattering medium. The second step is Fresnel diffraction from the scattering medium to the imaging plane. The light field after the scattering medium $dU'_s$ is equal to the one before multiplied by the transmission matrix of the scattering medium as, $dU'_s = dU'_s \cdot TM$, where the phase mask $TM(x_s, y_s)$ applied in the simulation is created from a random phase matrix convoluted with a 2D Gaussian function and thus, determining the frequencies of its spatial fluctuations[23].

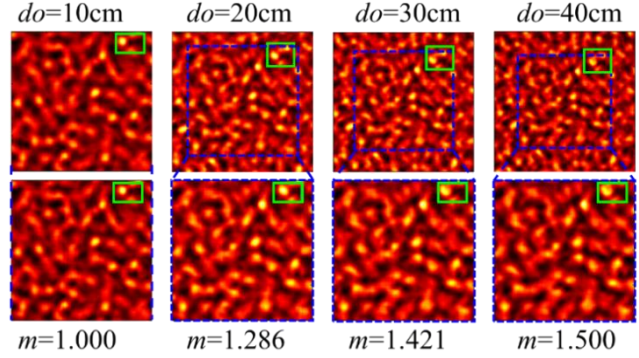

**Figure 2.** The speckles on the imaging plane derived from a pinhole located on different object planes. The first row is the original calculated results with green rectangles selecting the region of interest and blue squares indicating the area similar to the left-most image. The second row shows the images that are $m \times m$ times the size of the area in the blue squares.

By setting the pinhole as a 50 μm diameter round disk and applying to standard Fresnel diffraction algorithms, where both S-FFT for Fresnel diffraction propagation and D-FFT for angular spectrum propagation are tested, the intensity of the light field on the imaging plane is worked out as shown in Fig. 2. Obvious shrinking of speckle pattern can be observed when $d_o$ of pinhole is increasing. A scaling factor of $m$ can prevent the speckle pattern from shrinking.

On the other hand, the size of the speckles (also known as resolution) has been considered as relating to the aperture angle that radiation from the scattering medium as[25],

$$\sigma \approx 1.2\lambda d_i/D = 0.61\lambda/NA , \qquad (7)$$

where $D$ is the diameter of the scattering medium and $NA$ is numerical aperture of the scattering lens. Here we have to emphasize that $\sigma$ is also related to the object distances (the input light field). When $d_o$ is translating from 10 cm to 40 cm ($d_i$=10 cm), the whole speckle pattern is shrinking. One of the brightest spot in green rectangle is shifting towards the image center. The related area (in blue squares) is zoomed in by a factor of $m = f/f'$ and cut out (on the 2$^{nd}$ row of Fig. 2). The zoomed speckles show the same appearance with the reference speckle pattern ($d_o$= 10cm), which is consistent with the theoretical result.

**Experiment**

An experimental setup for imaging through a scattering medium is presented in Fig. 1. A physical pinhole illuminating by an incoherent light source (1W green LED source by Daheng Optics, GCI-060403) is placed on the optical axis of the imaging system. Its diffractive light projects onto a standard scattering medium (Newport 5° circular light shaping diffuser) and diffuses. A monochrome CCD (Basler ACA2040-90UM) is applied to capture the diffused light, where $d_i$ =7.5 cm. By translating the pinhole along the *z* direction from $d_o$ =11 cm to $d_o$ =24 cm with an interval of 0.5 cm, a series of *PSFs* is recoded (obviously zooming effect on the speckles can be seen the supplementary movie).

By selecting PSF of $d_o$ =15 cm as the reference, deconvolution of different PSFs with the reference PSF are executed. The results are shown in the 1$^{st}$ row of Fig. 3, which represent the reconstructed image of the pinhole in the corresponding object plane. The reconstructed images appear defocused and blur when their object planes are not exactly on the central object plane. The peak of the correlated images rapidly declines as the pinhole deviates from the central position. Its value drops down to 0.5 when the pinhole is translated to *z*~ ±3.6 mm, as shown in the Fig. 3(a). Hence the original DOF of deconvolution imaging technique is 7.2 mm (FWHM of Gaussian fit of Fig. 3(a)).

By rescaling the selected PSFs according to equation (6), the correlated images remain 'focused' over large distances with peak values declining slowly. Hence the axial range of retrieved imaging is improved to be 36.6 mm (FWHM of Gaussian fit of Fig. 3(b)), which is 5 times as large as the original DOF (shown in Fig. 3(d)).

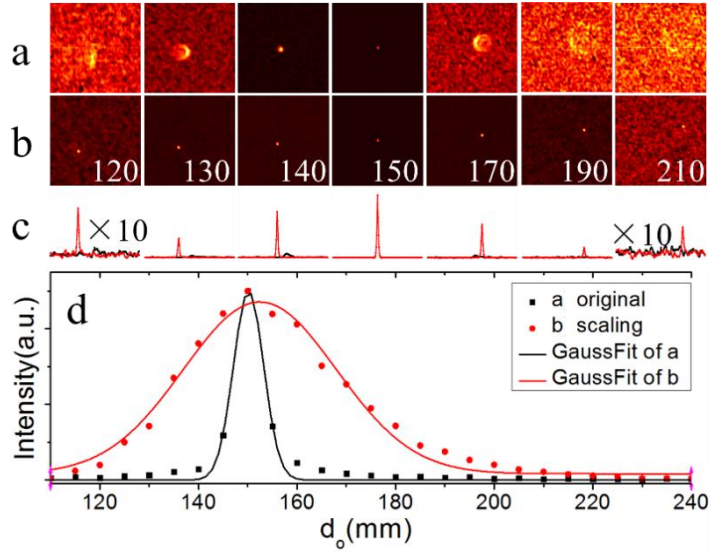

**Figure 3.** Deconvolution between PSFs from different $d_o$ with a selected PSF from central object plane ($d_o$=15cm). The selected PSF in (a) is the original one and in (b) are virtual PSFs after scaling, the inserted number indicates different $d_o$ in millimeter. (c) The cross section through the peaks of (a) black line and (b) red line. (d) Peak values of the intensity correlations.

For image restoration, the pinhole is replaced with unknown tested objects after the reference PSF ($d_o$=15 cm) is captured. The tested object (letter mask of 'H', about 1 mm tall) is located on different object distances (12~21 cm, step of 0.5 cm) once at a time and their speckles are captured separately. The reconstructed images are calculated by a deconvolution algorithm[4] with the original reference PSF. As shown in Fig. 4(a), the image of 'H' is resolvable only when the object locates within the original DOF range. By rescaling the reference PSF according to equation (6) (shown in Fig. 4(b)), the imaging of 'H' can be resolved clearly from $d'_o$ =12 cm to 19 cm. This range is even larger than the improved DOF (36.6 mm) because the rescaled PSFs are more relevant to the real PSFs of their object planes (at $d'_o$). Their correlation images would be similar to Fig. 3(b) and remain 'focused' over larger distances than the improved DOF range.

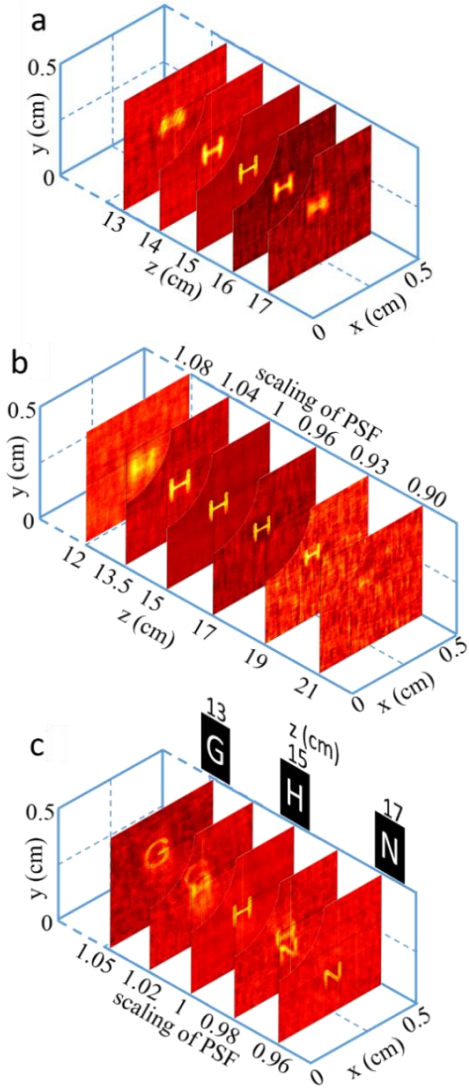

**Figure 4.** Image restoration with (a) a single PSF, and with (b) (c) its derived virtual PSFs. Objects in (a) (b) is a letter 'H' and in (c) are 3 letters located at different object plane simultaneously. The top left corners of each restored images are setting transparent for better view of the blocked layers.

When DOF of the imaging system is extended to a sufficiently large range, 3D imaging would naturally be achieved. Three different letter masks of 'G', 'H' and 'N' are placed on three separated object planes, as shown in Fig. 4(c). Their object distances are 13 cm, 15cm and 17cm respectively. Letters 'G' and 'N' are obviously located beyond the original DOF range with the only known PSF at $d_o$ =15 cm. The speckle pattern of these three letters is captured by the CCD. By rescaling the size of the reference PSF and correlating it with the objects speckle, images of the letter masks are restored separately with different scaling factor.

Furthermore, the proposed method can be used to determinate the location of a tested object. When the object distance is unknown, its image can be restored with an altering scaling factor searched by an adaptive algorithm[26]. The right scaling factor results in a restored image with highest contrast. And finally, the corresponding object distance is calculated according to equation (6).

**Discussion**

The relationship between the PSFs of scattering media from different reference points can be essentially derived from the angular ME, which states that the speckle pattern through a scattering medium is still highly correlated but shifted when the incident wavefront tilts slightly. Considering a pinhole moves towards the scattering medium along *z* axis direction from the plane at infinity, the wavevector from the pinhole to the same area of the scattering medium will tilt towards the vertical direction. Hence, its intrinsic isoplanatism on the imaging plane will expand outward but still 'correlated' (size should be rescaled) following the angular ME. However, the max tilt angle before 'uncorrelated' is limited by ME angle, so there still is a limitation of $\Delta d_o = d'_o - d_o$, namely improved DOF.

The original resolution (equation (7)) of an imaging system based on scattering medium is derived from plane wave incident (red line), in which cases $d_o$ is infinite and $f_\infty = d_i$. In the process of imaging, the object distance should no longer be infinite, the size of the speckle pattern as well as the resolution would expand with *m* times ($m = f_\infty / f$), as presented in Fig. 2. Equation (7) becomes,

$$\sigma = m \frac{1.2 \lambda d_i}{D} = \frac{1.2 \lambda d_i^2}{Df} = \frac{0.61 \lambda}{NA'}, \tag{8}$$

where $NA' = \dfrac{Df}{2d_i^2}$ is the equivalent *NA* for the imaging processing with a scaling lens.

According to Ref. 17 and Ref. 27, the nominal DOF[17, 27], which depicted the axial correlation of the speckles in the imaging plane, equals to about $6.7\lambda(z/D)^2$. In our scene, it should be modified with *NA'* and be deviated to the objective plane. It turns out to be $DOF=6.7\lambda/M^2NA'^2=7.3mm$, which is closed to the original DOF (7.2 mm) in our experiment, where $M=d_i/d_o$ is the transverse magnification.

The improved DOF is calculated with deconvolution (or correlation) processing, so it should be limited by the axial FOV. The form factor *F* in equation (1) indicates the transverse FOV. It is commonly defined as $FOV \sim \lambda d_o/\pi L$ and measured by correlating the speckle patterns of a transverse shifting point. Katz et al.[28] have deduced the axial FOV of a scattering lens compensated with a spatial light modulator from its transverse FOV. A discussion on calculating the axial FOV of a scattering medium with our rescaling method can be found in the supplementary. The maximum axial FOV is estimated as $\Delta z_{FOV}=1.4d_o FOV/D \sim 61mm$, which is larger than our improved DOF 36.6 mm. Several factors in the experiment degrade the correlation of the speckles and shrink the improved DOF, including low speckle contrast under broadband spectrum illumination, transverse deviation of the object (Fig. 3(b)), low sampling for the small size speckle, discretization error when rescaling and etc. Light source generated by laser passing through high speed rotating diffuser and a digital camera with smaller pixel pitch will make the DOF of our method closer to the axial FOV limit. One simpler way to enlarge the DOF is adding a diaphragm before the scattering medium. Another method is to decrease the angular difference by inserting lens[4]. Although the improved DOF is still limited by the axial FOV, the dependence of speckle size on the objective distance in equation (8) are valid (expects the complicated "near field speckles" regime). It is possible to separate or extract the speckle patterns of the objects placing beyond axial FOV and to realize a 3D Non-invasive single-shot imaging. Our method can also improve the imaging quality of lensless cameras[29] (using pseudo-random phase mask with convex bumps) and of lensless microscopy[30] (using a scatter-plate with a thin scattering layer) and adapt them to more complex diffusers. Solid-state samples with different strength of scattering (commercial standard scattering samples from 5deg to 30deg), strong forward biological scattering samples (1mm chicken tissues) are tested with the proposed method. The results confirm that the DOF can be enlarged for more samples whose thicknesses are larger than several mean free paths[24].

## Conclusion

The correlation of the scattering light in axial direction is studied and exploited. We reveal a rigorous relationship between the PSFs of a thin scattering medium from different reference points. By adjusting the scale of one PSF, other PSFs from different object planes can be deduced and obtained. Theoretical derivation, prove-of-concept simulation and experiment for analyzing this relationship are demonstrated with good quality and high fidelity. Experimental results show an approximate 5 times of improvement in depth resolved ability compared to the original method without PSF operation. 3D objects located separately beyond the original DOF is reconstructed through a thin scattering medium slide by slide with the proposed single PSF deconvolution technique. For a known reference object place, our technique can be even made use of detecting the distance in axial direction through a thin scattering medium.

## Methods

The numerical simulation method includes two steps of Fresnel diffraction, as shown in Fig. 1, the first step is from the pinhole to the plane in front of the scattering medium. The second one is from the scattering medium to the imaging plane. The phase mask $TM(x_s, y_s)$ applied in the simulation is created from a random phase matrix convoluted with a 2D Gaussian function and thus, determining the frequencies of its spatial fluctuations[23]. By setting the pinhole as a 50 μm diameter round disk and applying to standard Fresnel diffraction algorithms, where both S-FFT for Fresnel diffraction propagation and D-FFT for angular spectrum propagation are tested, the intensity of the light field on the imaging plane is shown in Fig. 2.

The experimental setup as shown in Fig. 1, A physical pinhole illuminating by an incoherent light source (1W green LED source by Daheng Optics, GCI-060403) is placed on the optical axis of the imaging system. Its diffractive light projects onto a standard scattering medium (Newport 5° circular light shaping diffuser) and diffuses. A monochrome CCD (Basler ACA2040-90UM) is applied to capture the diffused light, where $d_i$ =7.5 cm. To reveal the physical relationship between the s PSFs from different object planes after a thin scattering medium, a pinhole is translating along the *z* direction from $d_o$ =11 cm to $d_o$ =24 cm with an interval of 0.5 cm, a series of *PSFs* is recoded. By selecting PSF of $d_o$ =15 cm as the reference, deconvolution of different PSFs with the reference PSF are executed, as shown in Fig.3.

For image restoration, the pinhole is replaced with unknown tested objects after the reference PSF ($d_o$=15 cm) is captured. The tested object (letter mask of 'H', about 1 mm tall) is located on different object distances (12~21 cm, step of 0.5 cm) once at a time and their speckles are captured separately. By rescaling the reference PSF according to equation (6) (shown in Fig. 4(b)), the imaging of 'H' can be resolved clearly from $d'_o$ =12 cm to 19 cm with an improved DOF (36.6 mm). When DOF of the imaging system is extended to a sufficiently large range, 3D imaging would naturally be achieved. Three different letter masks of 'G', 'H' and 'N' are placed on three separated object planes, as shown in Fig. 4(c). Their object distances are 13 cm, 15cm and 17cm respectively. Letters 'G' and 'N' are obviously located beyond the original DOF range with the only known PSF at $d_o$ =15 cm. The speckle pattern of these three letters is captured by the CCD. By rescaling the size of the reference PSF and correlating it with the objects speckle, images of the letter masks are restored separately with different scaling factor.

## Acknowledgements

This work is supported by National Natural Science Foundation of China (61575223, 11534017, 61705035 & 61475038), State Key Laboratory of Optoelectronic Materials and Technologies (Sun Yat-sen Unversity), STU Scientific Research Foundation for Talents.

## Author contributions

All the authors discussed, interpreted the results and conceived the theoretical framework. Xie and He conceived the initial concept and wrote the paper. Zhuang and Liang designed and performed the experiments. Zhuang, Xu and Liu designed the simulations. Zhou provided the whole experimental platform. All authors reviewed the manuscript.

## Additional information

The authors declare no competing financial interests.